\documentstyle[twoside,epsfig,psfig]{article} 
 
\input ibvs2.sty 
 
\begin{document} 
\IBVShead{6087}{20 Dec 2013} 
 
\IBVStitle{Photometry of the progenitor of Nova Del 2013 (V339 Del)}
\vskip -0.3 cm
\IBVStitle{and calibration of a deep BVRI photometric}
\vskip -0.3 cm
\IBVStitle{comparison sequence}
\IBVStitle{} 
  
\IBVSauth{Ulisse Munari$^{1,2}$, Arne Henden$^3$} 

\IBVSinst{INAF Osservatorio Astronomico di Padova, Sede di Asiago, 
          I-36032 Asiago (VI), Italy} 
\IBVSinst{ANS Collaboration, c/o Astronomical Observatory, 
          I-36012 Asiago (VI), Italy} 
\IBVSinst{AAVSO, 49 Bay State Rd. Cambridge, MA 02138, USA}

\IBVStyp{Nova} 
\IBVSkey{photometry} 
\IBVSabs{The Asiago plate archive has been searched for old plates covering
the region of the sky containing Nova Del 2013 (V339 Del).  The brightness
of the progenitor was measured against a deep BVRI photometric sequence that
we calibrated on purpose.  The mean brightness of the progenitor on Asiago
plates is <B>=17.27 and <V>=17.6, for a mean color (B-V)=-0.33.  The
recorded total amplitude of variation in B band is 0.9 mag.  Color and
variability are in agreement with a progenitor dominated by the emission
from an accretion disc.  The progenitor was marginally detected also by the
APASS all sky survey on April 2012.  We have stacked the CCD images from
three individual visits and measured the progenitor at B=17.33+/-0.09 and V
=17.06+/-0.10 mag.}

\begintext 

Nova Del 2013 (=V339 Del) was discovered on 2013 Aug 14.584 UT by K. 
Itagaki when it was already shining at unfiltered 6.8 magnitude (cf.  CBET
3628).  The observation by Denisenko et al.  (cf CBET 3628) reporting the
nova still in quiescence at $\sim$17.1 mag on Aug 13.998 UT (14 hours before
the discovery), would indicate a very fast rise to maximum.  The photometric
evolution of the nova during the optically thick phase has been presented by
Munari et al.  (2013a).

The nova appeared at a position coincident with the blue star USNO B-1
1107-0509795, reported at $B$$\sim$17.20, $R_{\rm C}$$\sim$17.45 on the first
Palomar Sky Survey plates (exposed on 7 July 1951), and at $B$$\sim$17.39,
$R_{\rm C}$$\sim$17.74 on the second Palomar Sky Survey plates (exposed on 18
July and 15 September 1990, respectively). The progenitor is bright enough
to have been recorded on patrol plates taken with the Asiago Schmidt
telescopes, in particular the 67/92 cm instrument. We searched the
Asiago plate archive and found 25 plates imaging the area of the sky where
the nova later appeared (16 in $B$ band, 7 in $V$, and 2 in $I_{\rm C}$).

Prior to measuring these plates, it was necessary to establish an accurate
and reliable photometric sequence around the nova. Given the huge interest
this nova has received because of its peak brightness$(V$=4.46, $B$=4.70), it
may be presumed that its remnant will be studied for long after the system
has returned to quiescence. To ensure proper comparison of the
pre-outburst photometric data with the post-outburst ones, a well calibrated
and common photometric sequence must be adopted. To this aim we obtained
deep exposures in $B$$V$$R_{\rm C}$$I_{\rm C}$ bands of the field
surrounding the nova on three separate photometric nights with
the TMO61 telescope (part of the AAVSOnet robotic network) in New Mexico (USA).
The data for each night were independently calibrated against the equatorial
Landolt (2009) standard stars, and the results averaged. 

From the common data set, we have extracted two sets of standards, presented
in Tables~1 and 2.  These standards are on the same photometric scale of the
brighter ones used by Munari et al.  (2013a) to derive the photometric
evolution of the nova around maximum and early decline.  The standards in
Table~1 are optimized for CCD observations and are grouped within 5 arcmin
of the nova.  They were selected at three different\\

\IBVSfig{12cm}{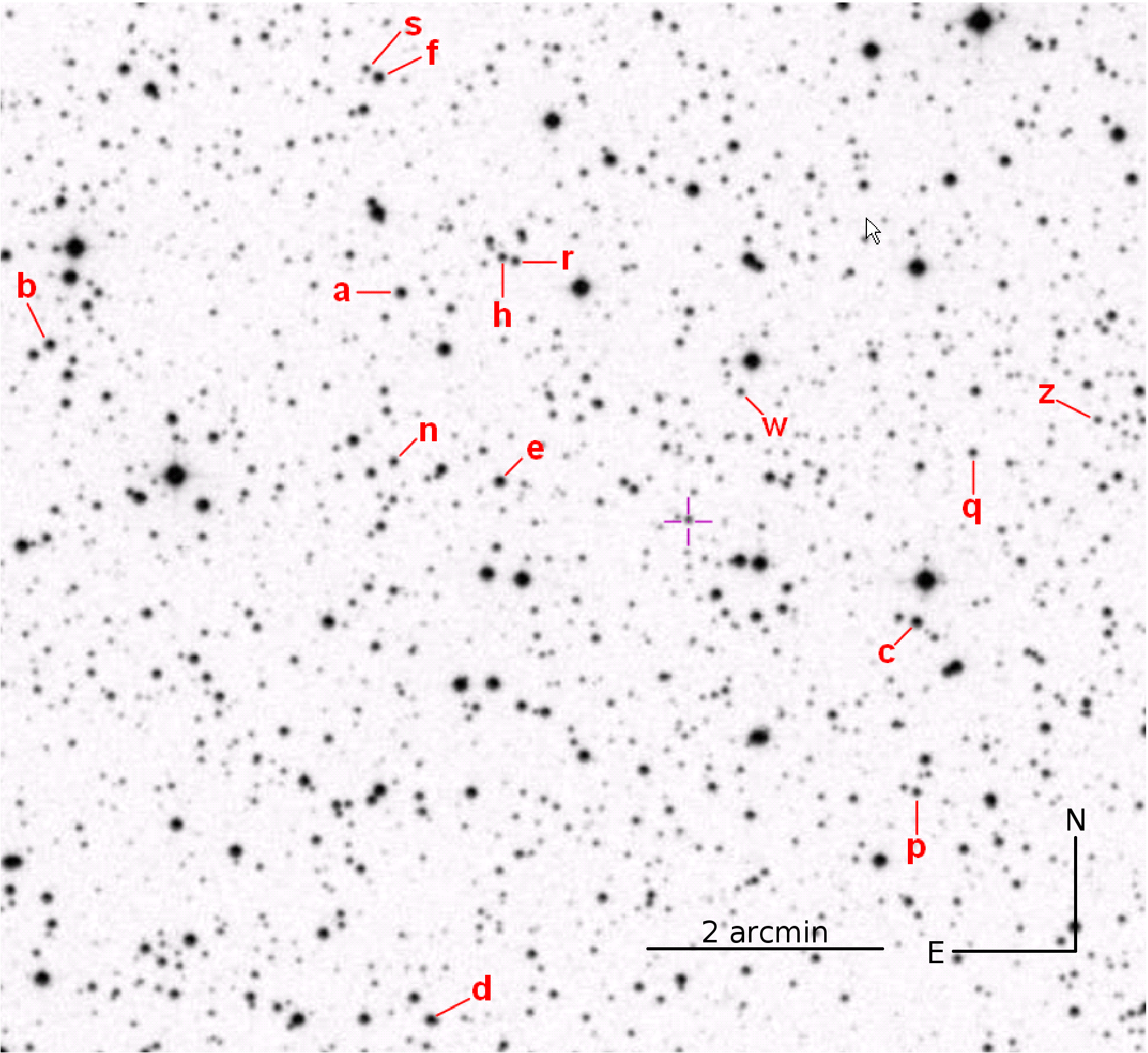}{Finding chart for the $B$$V$$R_{\rm C}$$I_{\rm C}$ 
photometric sequence in Table~1, optimized for CCD observations of Nova Del
2013 during the advanced decline and after its return to quiescence.}

\IBVSfig{6cm}{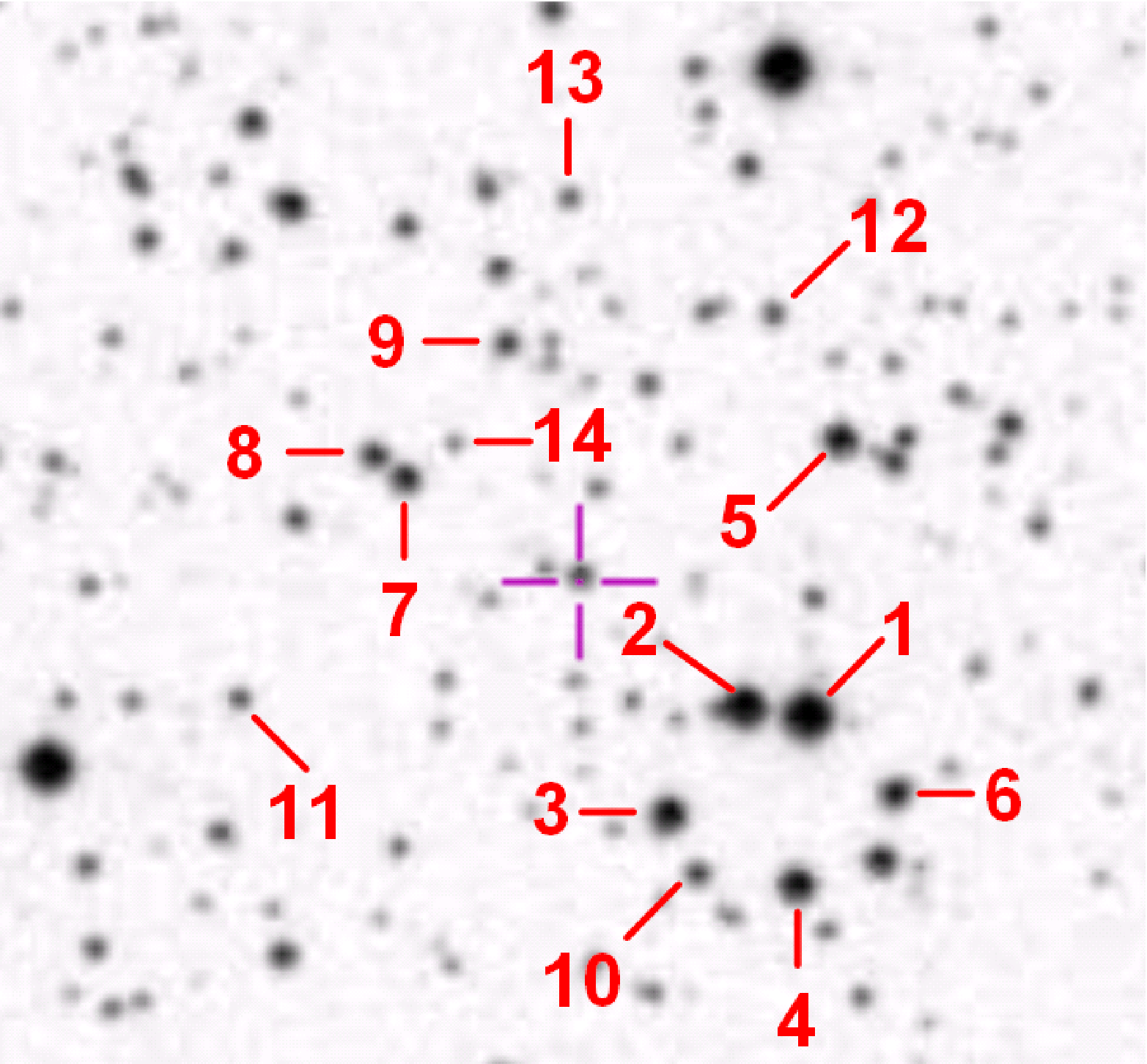}{Finding chart for the $B$$V$$R_{\rm C}$$I_{\rm C}$ 
photometric sequence in Table~2, optimized for measurement of the
progenitor on archive photographic plates. The chart is 3.0 arcmin wide with
North to the top and East to the left.} 

\clearpage
\begin{table}[!Ht]
  \caption{$B$$V$$R_{\rm C}$$I_{\rm C}$ photometric sequence (plotted in
   Figure~1), optimized for CCD observations of Nova Del 2013 during the
   advanced decline and after its return to quiescence.}
  \centerline{\psfig{width=15cm,file=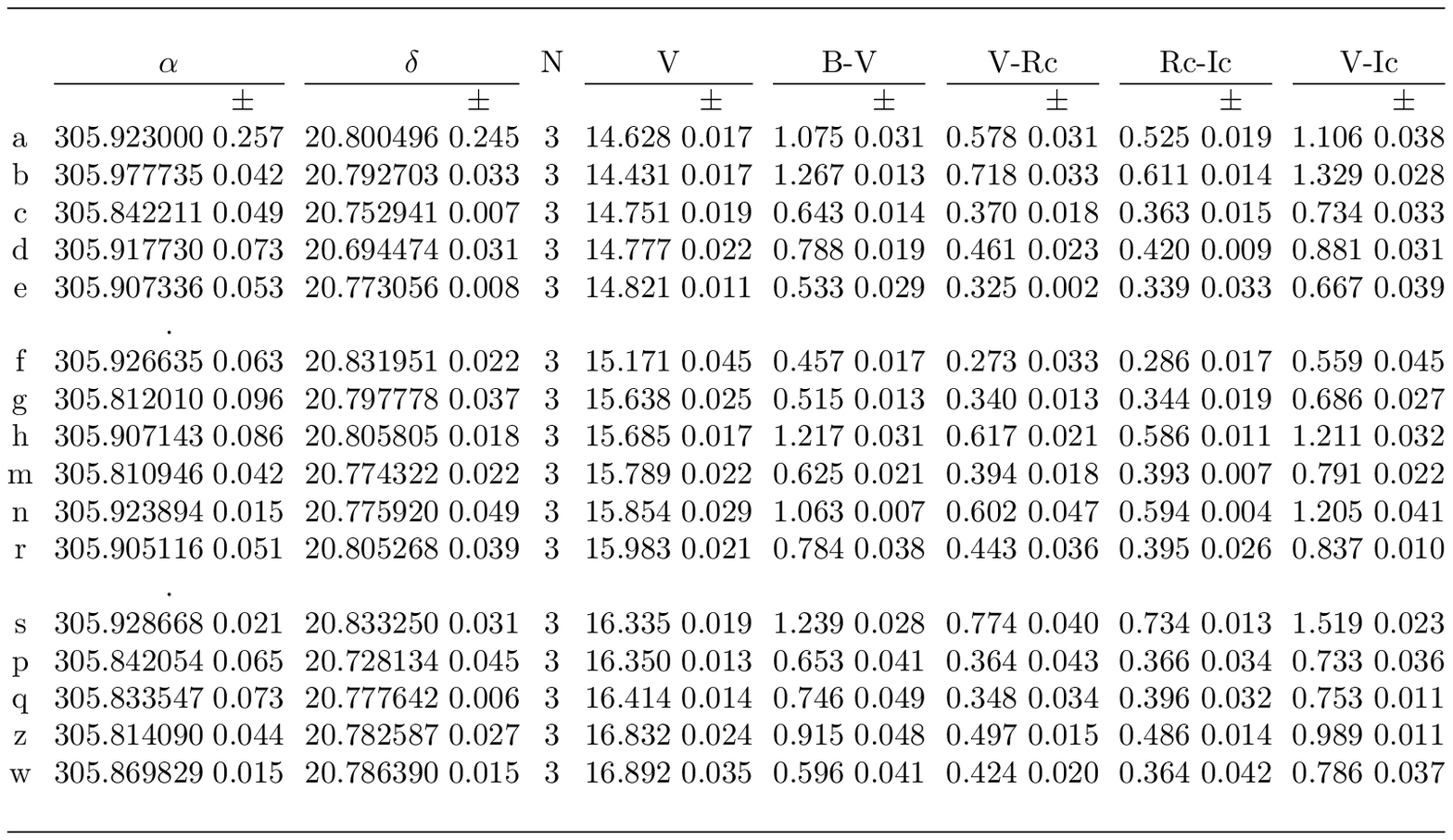}}
\end{table}

\begin{table}[!Ht]
  \caption{$B$$V$$R_{\rm C}$$I_{\rm C}$ photometric sequence (plotted in
  Figure~2), optimized for measurement of the progenitor on archive
  photographic plates.}
  \centerline{\psfig{width=15cm,file=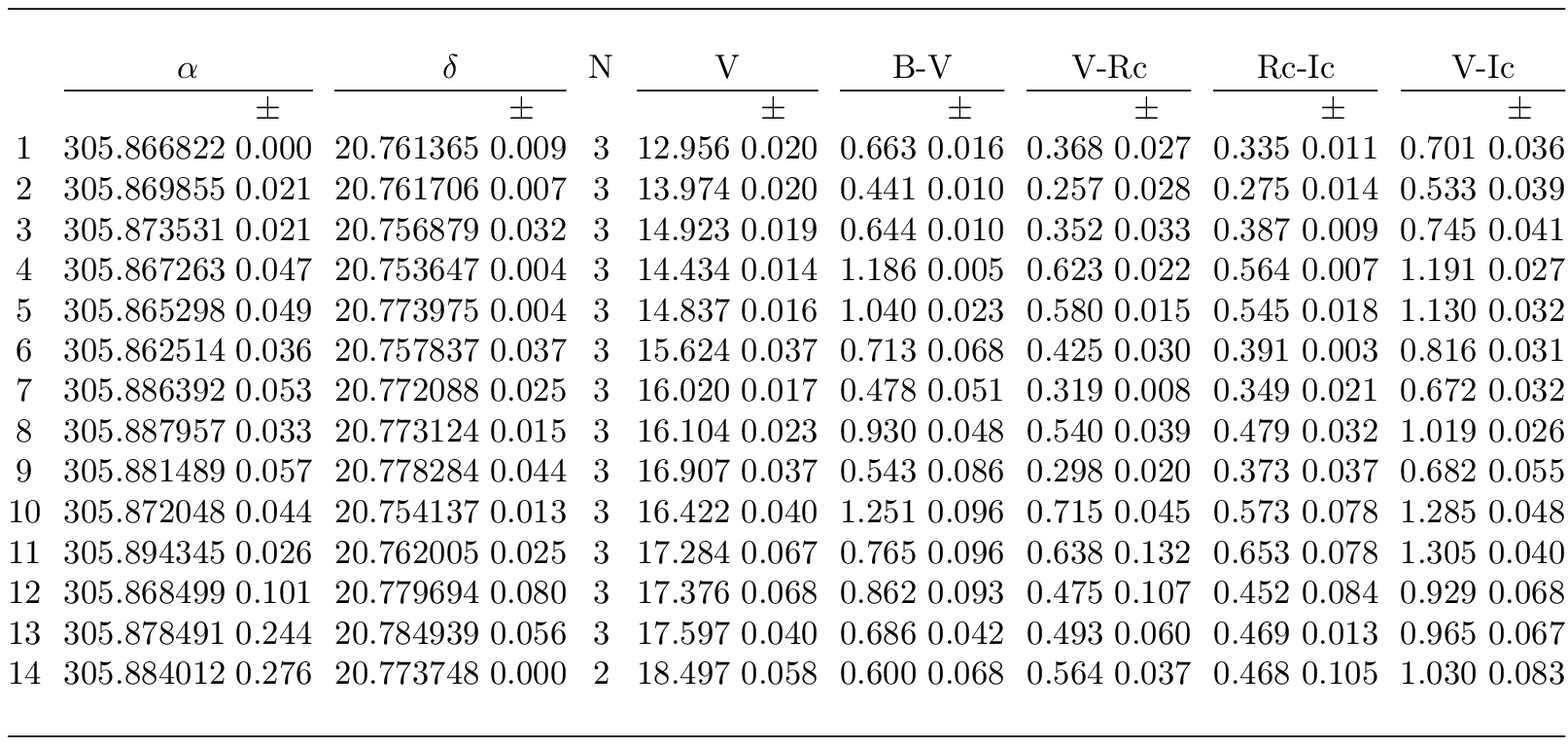}}
\end{table}

\noindent
magnitude levels ($V$$\sim$14.6, 15.6, and 16.6 mag) to support photometric
investigation of both the advanced decline and the following return to
quiescence of the nova.  At each of the three different magnitude levels, at
least five standards are provided that cover a broad range of colors so to
allow the calibration of color equations to transform the measurements from
the local to the standard system.  The standards in Table~2 are instead
optimized to derive the magnitude of the progenitor on old photographic
plates, most of which were exposed in blue light or in $B$ band.  They are
grouped within 1 arcmin of the nova.

\begin{table}[!Ht]
  \caption{The brightness of the progenitor of Nova Del 2013 as measured on
           Asiago photographic plates.}
  \centerline{\psfig{width=13cm,file=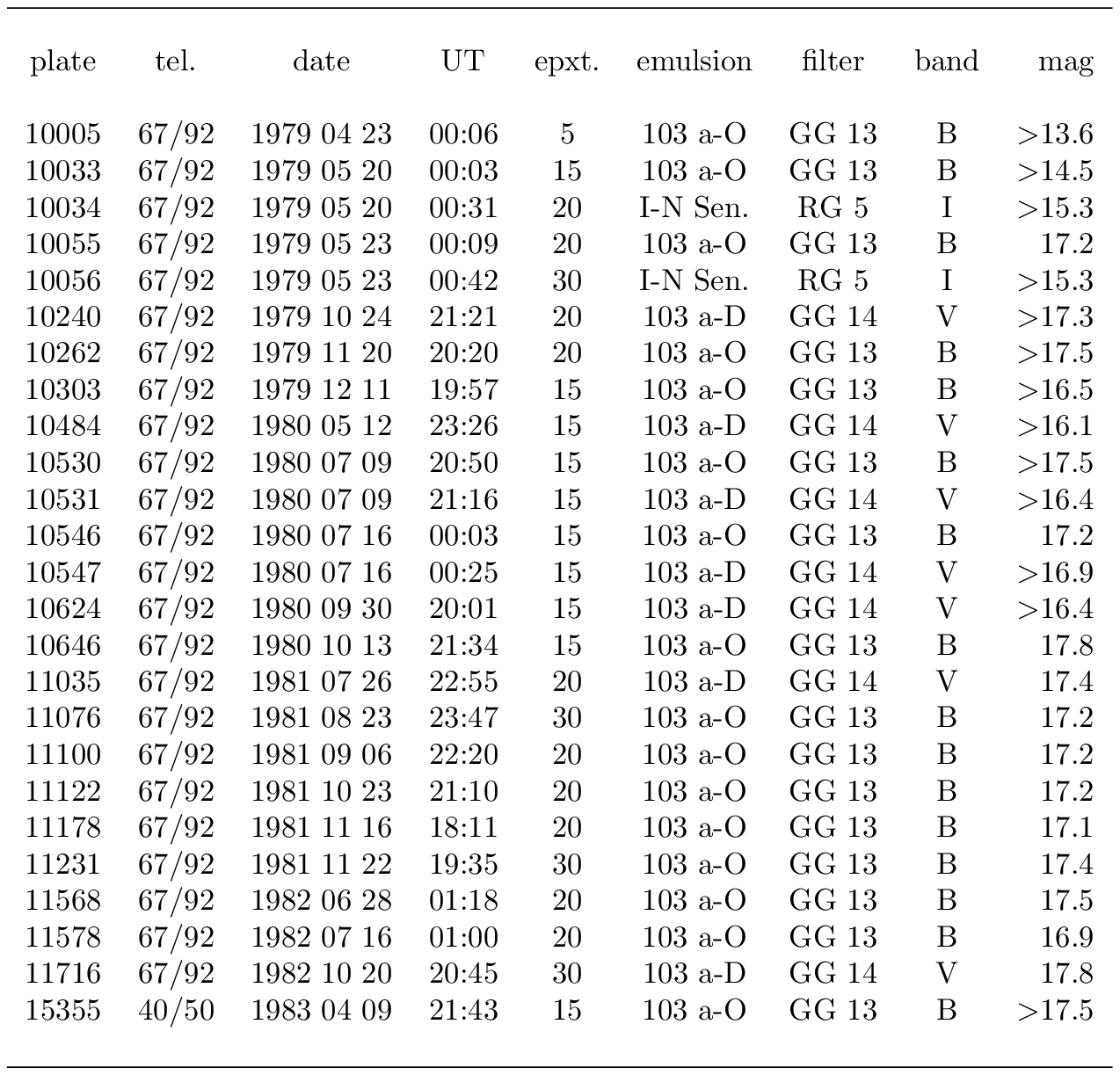}}
\end{table}

The magnitude of the progenitor of Nova Del 2013 on Asiago photographic
plates was estimated by eye through a high quality Zeiss microscope.  The
plates were independently re-measured a few days later and the results
were found to be the same within 0.1 mag, which is therefore taken as the
error associated to the measurements.  The results are presented in Table~3. 
The mean brightness of the progenitor on these plates is $<B>$=17.27 and
$<V>$=17.6, for a mean color ($B$$-$$V$)=$-$0.33.  The recorded total
amplitude of variation in $B$ band is 0.9 mag.  Color and variability are in
agreement with a progenitor dominated by the emission from an accretion
disc.  The reddening toward Nova Del 2013 is low ($E_{B-V}$=0.18, e.g. 
Munari et al.  2013b) given its high galactic latitude
($b$=$-$9$^{\circ}_{.}4$).  Brightness level and color are in excellent
agreement with the USNO-B1 values from Palomar surveys 1 and 2, arguing for
a long term stability before the 2013 eruption.

The progenitor was marginally detected also by the APASS all sky survey,
when it visited the field on 2012 April 21, 24 and 25, thus about 18 months
before the nova eruption.  We have stacked the CCD images from these three
visits and measured the progenitor at $B$=17.33$\pm0.09$ and $V$=17.06$\pm0.10$.

\references 

Landolt, A.~U. 2009, AJ 137, 4186

Munari, U. et al. 2013a, IBVS 6080

Munari, U. et al. 2013b, ATel 5297

\endreferences 
\clearpage

\end{document}